\documentclass[a4paper,fleqn,usenatbib]{stasclass}

\usepackage{newtxtext,newtxmath}

\usepackage[T1]{fontenc}
\usepackage{ae,aecompl}

\usepackage{graphicx}	
\usepackage{amsmath}	
\usepackage{amssymb}	

\title[On the temperature of the solar wind]{On the Temperature of the Solar Wind}

\author[Boldyrev {et al.}]{
Stanislav Boldyrev,$^{1, 2}$\thanks{E-mail: boldyrev@wisc.edu}
Cary Forest,$^{1}$
and Jan Egedal$^{1}$
\\
$^{1}$Department of Physics, University of Wisconsin -- Madison, 1150 University Avenue, Madison, WI 53706, USA\\
$^{2}$Space Science Institute, Boulder, CO 80301, USA
}

\date{}

\pubyear{2020}

\begin{document}
\label{firstpage}
\pagerange{\pageref{firstpage}--\pageref{lastpage}}
\maketitle

\begin{abstract}
Solar wind provides an example of a weakly collisional plasma expanding from a thermal source in the presence of spatially diverging magnetic field lines. Observations show that in the inner heliosphere, the electron temperature declines with the distance approximately as $T_{e}(r)\sim r^{-0.3} \dots r^{-0.7}$, which is significantly slower than the adiabatic expansion law $ \sim r^{-4/3}$. Motivated by such observations, we propose a kinetic theory that addresses the non-adiabatic evolution of a nearly collisionless plasma expanding from a central thermal source. We  concentrate on the dynamics of energetic electrons propagating along a radially diverging magnetic flux tube. Due to conservation of their magnetic moments, the electrons form a beam collimated along the magnetic field lines. Due to weak energy exchange with the background plasma, the beam population slowly loses its energy and heats the background plasma. We propose that no matter how weak the collisions are, at large enough distances from the source a universal regime of expansion is established where the electron temperature declines as $T_e(r)\propto r^{-2/5}$. This is close to the observed scaling of the solar wind temperature in the inner heliosphere. Our first-principle kinetic derivation  may thus provide an explanation for the slower-than-adiabatic temperature decline in the solar wind.  More broadly, it may be useful for describing magnetized winds from G-type stars. 
\end{abstract}




\section{Introduction}
\label{introduction}
The solar wind is formed by a magnetized and nearly collisionless plasma streaming from the hot solar corona. Observations demonstrate that as the wind expands with the heliospheric distance, its temperature gradually declines. (This is true below the distances of about $10-20$ astronomical units (AU). At larger distances, the temperature, in fact, increases with the distance due to the heating provided by pickup ions~\cite[e.g.,][]{richardson2003,cranmer2009,koehnlein96,bale2016}). The solar wind, however, is not cooled down as fast as one would expect based on the adiabatic expansion law. While at the corona the temperature is on the order of hundred electron-volt,
it decreases to about $10$~eV at $1$~AU. If the plasma expansion were adiabatic, the temperature at $1$~AU would be order of magnitude smaller than observed, which indicates the presence of significant non-adiabatic heating. The mechanism of solar wind heating is a long-standing puzzle of space plasma physics. It is crucial for understanding of the characteristics of the solar wind, which have been investigated by spacecraft over a range of heliospheric distances and latitudes (e.g., Helios, Ulysses, Voyager, Wind missions), and will be probed at the distances unprecedentedly close to the sun by the new Parker Solar Probe mission. 
 
Another puzzle is provided by the observational fact that the ion and electron temperatures are different in the solar corona. For instance, according to the multifrequency radio imaging measurements \cite[e.g.,][]{mercier2015}, the electron temperature at the solar corona is about $3.5$ times smaller than the ion  one; see also \cite{noci2003,landi2007,landi2009}. In addition, the temperature declines with the heliospheric distance somewhat differently in the slow and fast solar wind. In the slow wind, the temperature profile in the inner heliosphere is characterized by an approximate power law $T_e(r)\propto r^{-0.5}\dots r^{-0.7}$, while in the fast and less collisional wind, the decline is slower $T_e(r)\propto r^{-0.3}\dots r^{-0.4}$, \cite[e.g.,][]{stverak15}. The latter difference is, however, not going to be essential for our consideration.

Several mechanisms may contribute to the non-adiabatic temperature profile of the solar wind. For instance, the plasma may be heated as a result of instabilities and turbulent fluctuations that extract energy from the streaming motion and convert it into kinetic energy of particles \cite[e.g.,][]{richardson2003,cranmer2007,cranmer2009,chen2016,vech2017,tang2018,bercic2019,verscharen2019,verscharen2019a,lopez2019,shaaban2019,vasko2019,RC2019}. The analysis of such local instabilities, however, does not allow for a definitive prediction of the global radial temperature profile.  In this work, we discuss a different (and complementary) heating mechanism related to the electron energy transport and energy deposition into background plasma governed by weak Coulomb collisions. The fast electrons streaming from the hot corona along the magnetic field lines, experience weak collisions, so they can transport energy to large heliospheric distances with relatively little attenuation. Due to conservation of their magnetic moment, such electrons form the so-called strahl,\footnote{We broadly define a strahl as a distribution function of streaming electrons collimated by magnetic field, no matter what the energy of the electrons is. In the solar-wind literature, it is also common to apply this term to the collimated electrons whose energy exceeds a certain threshold, typically a few thermal energies of the core electron distribution.} an electron beam collimated around the direction of the background magnetic-field lines \cite[e.g.,][]{feldman75,pilipp87,scudderolbert79,pierrard2016,horaites15,horaites18a,horaites18b,horaites2019,boldyrev19}. Due to weak Coulomb collisions, the strahl slowly loses its energy and heats the plasma. 

Heat conduction by electrons was considered previously in the models of solar wind heating \cite[e.g.,][]{pilipp87,cranmer2009,stverak15}. We, however, note that heat conduction is a collisional process, while the electron strahl is a nearly collisionless phenomenon. The processes by which the kinetic energy is extracted from the strahl and deposited into the thermal energy of the bulk plasma are therefore not described by the standard theory of heat conduction. Indeed, the \cite{spitzerharm53} theory of heat conduction  requires that the mean-free path of the electrons be significantly smaller than the temperature-gradient scale. The ratio of the two scales, the Knudsen number $Kn=\lambda_{\rm mfp}/L_T$, should satisfy $Kn\lesssim 0.01$ for the collisional theory to hold \cite[e.g.,][]{gurevichistomin79}. 
In our description of a nearly collisionless stellar wind, we, therefore, need to consider the case of large (but not infinitely large) Knudsen numbers. 


It is important to note that the electron thermal velocity is much lager than the velocity of the solar wind so the latter can be neglected in our discussion, and the background plasma and the magnetic-field configuration can be considered stationary.  In our theory, we also make another simplification -- we assume a magnetic flux tube where the magnetic lines diverge radially from the source. This is a good approximation for the inner heliosphere, where the toroidal component of the magnetic field, $B_\phi$, is generally smaller than the radial one,~$B_r$. This will suffice for our consideration. 

As mentioned above, due to the conservation of their magnetic moments, the electrons streaming from the sun into the region of weakening magnetic field, tend to form a progressively narrowing beam around the magnetic-field lines. This process is rather efficient already at distances very close to the sun. Weak Coulomb collisions, on the other hand, broaden the beam, so that a narrow collimation angle $\theta$ is established.


We argue that plasma 
expansion leads to preferential cooling of the source electrons so that in a steady state, the electron temperature of the source should be smaller than the ion temperature. This general result is consistent with observations of the ion and electron temperatures in the solar corona. We also demonstrate that no matter how weak the collisions are, if the distance from the source is large enough, so that the condition $\sin^2\theta \lesssim r/\lambda_{\rm mfp}\ll 1$ is satisfied (below we will refer to such distances as the {\em far zone}), a universal heating  regime is established with the electron temperature declining as $T_e(r)\propto r^{-2/5}$. This result is close to the temperature profile observed in the solar wind.

Note that although collisions play an essential role in our analysis, the plasma is still formally collisionless,  since we assume $r/\lambda_{\rm mfp}\ll 1$. Some ideas of our analysis stem from the theory developed earlier for expansion regions of plasma mirror machines \cite[][]{konkashbaev1978,mirnov1979,ryutov2005}, where, however, collisions were so rare, that the opposite inequality $r/\lambda_{\rm mfp} \ll \sin^2\theta \ll 1$ was satisfied and as a result, a  different temperature profile, $T(r)\propto r^{-4/3}$, was predicted (we will refer to such distances as the {\em near zone}). 
In what follows, we first establish the analogy of these previous results with the problem of solar wind heating,  then present our solution for the far zone and apply it to the problem of solar wind temperature.

\section*{The electron distribution function}
\label{evdf}
In this section we consider a plasma expanding from a hot localized source along spatially diverging magnetic-field lines. We will assume that the ions and the electrons have Maxwellian distributions at the source region where their dynamics are collisional, but that the plasma outflow is nearly collisionless in the magnetic-field expansion region outside the source. A similar situation has been studied analytically in relation to particle confinement in mirror machines \cite[e.g.,][]{mirnov1979,post1987,ryutov2005,ivanov2013,ivanov2017}, where the extent of the expansion region, that is, the region between the throat of the mirror machine and the wall of the end tank, is many orders of magnitude smaller than the electron Coulomb mean-free path. An important result of the theory is that in a steady state, the total electron distribution function in the expansion region is not Maxwellian, rather, it has two distinct components.  The first is a tenuous collisionless beam of energetic electrons propagating away from the source. The second is a collisional and nearly isotropic ``core'' consisting of the so-called trapped particles (see also the earlier analysis in \citet{perkins1973}). 

In order to understand this result, let us assume, somewhat idealistically, that the plasma is collisional close to the source, but becomes collisionless beyond certain distance $r_0$.  For the solar wind, we may estimate $r_0\sim 5 -10 $ solar radii. Since the electrons have larger thermal velocities, they will leave the source faster so that the source will acquire a positive ``ambipolar'' potential with respect to infinity.  We will measure the potential with respect to its value at the source, $\phi(r_0)=0$, and denote the potential very far from the source as~$\phi_\infty$. This potential barrier will return most of the streaming electrons, except for the very energetic ones, back to the source.

The potential barrier can be found from equating the electron and ion fluxes leaving the source, taking into account that only the electrons whose energies exceed the potential barrier~$e\phi_\infty$ will eventually run away. Assuming, for simplicity, that the source is spherical and that the particle distribution at the boundary of the source is Maxwellian, we require that the fluxes of the ions and electrons through the boundary are the same, 
\begin{eqnarray}
\int f_{i,0}(v)v_\|\theta\left(v_\|\right) \,d^3v=\nonumber \\
=\int f_{e, 0}(v)v_\|\theta\left(v^2-2e\phi_\infty/m_e\right)\theta\left(v_\|\right)d^3v, 
\end{eqnarray}
where $f_{i,0}(v)$ and $f_{e, 0}(v)$ are the Maxwellian distributions, $T_{i,0}$ and $T_{e, 0}$ the corresponding temperatures, and $\theta$ is the Heaviside step function. As a result, we get:
\begin{eqnarray}\label{eq1}
\left(\frac{e\phi_{\infty}}{T_{e, 0}}+1\right)\exp\left\{-\frac{e\phi_{\infty}}{T_{e, 0}} \right\}
=\left(\frac{T_{i, 0}}{T_{e, 0}} \right)^{1/2}\left(\frac{m_e}{m_i}\right)^{1/2}.
\end{eqnarray}
It may seem that this equation allows one to find the required potential~$\phi_\infty$. This is, however, not the case. Indeed, if one assumes that $T_{e, 0}=T_{i, 0}$, the escaping electrons will carry away larger energy per unit time than the ions. Unless there is a heat source preferentially heating the electrons, their temperature will drop compared to the ion one. The electron temperature is, therefore, itself a parameter that needs to be found. 

If the ions and the electrons are heated at the {\em same} rate, then by equating their energy fluxes through the boundary of the source region, we obtain in a steady state
\begin{eqnarray}
\int f_{i,0}(v)m_iv^2 v_\|\theta\left(v_\|\right) \,d^3v=\nonumber \\
=\int f_{e, 0}(v)m_ev^2 v_\|\theta\left(v^2-2e\phi_\infty/m_e\right)\theta\left(v_\|\right)d^3v.
\end{eqnarray}
As a result, we get:
\begin{eqnarray}\label{eq2}
\left[\frac{1}{2}\left(\frac{e\phi_{\infty}}{T_{e, 0}} \right)^2 +\frac{e\phi_{\infty}}{T_{e, 0}}+1\right]\exp\left\{-\frac{e\phi_{\infty}}{T_{e, 0}} \right\}=
\nonumber \\
=\left(\frac{T_{i, 0}}{T_{e, 0}} \right)^{3/2}\left(\frac{m_e}{m_i}\right)^{1/2}. 
\end{eqnarray}
From equations (\ref{eq1}) and~(\ref{eq2}), one can derive the resulting potential barrier and the electron temperature. For instance, assuming a hydrogen plasma, one finds that $e\phi_\infty  \approx 5 T_{e, 0}$, and that the electrons in the source region should be essentially {\em colder} than the ions, $T_{e, 0}\approx T_{i, 0}/3$, see also~\cite{ryutov2005} .  
Note that this effect alone is consistent with, and may provide an explanation for, the observational inequality of the ion and electron temperatures in the solar corona \cite[e.g.,][]{noci2003,landi2007,landi2009,mercier2015}.

If, however, the electrons and the ions are heated differently, this estimate needs to be modified. For instance, in some solar corona models it is proposed that the ions are heated more efficiently, so that  $T_{e, 0}/T_{i, 0} \approx 0.1$ \cite[e.g.,][]{chandran10}. If this ratio is substituted into Eq.~(\ref{eq1}), we obtain $e\phi_\infty\approx 4.2 T_{e, 0}$. Both cases demonstrate that the electron temperature has to be {\em smaller} than the ion one, and that the potential barrier $e\phi_\infty$ is several times larger than~$T_{e, 0}$. 

\begin{figure}
\centerline{\includegraphics[width=\columnwidth]{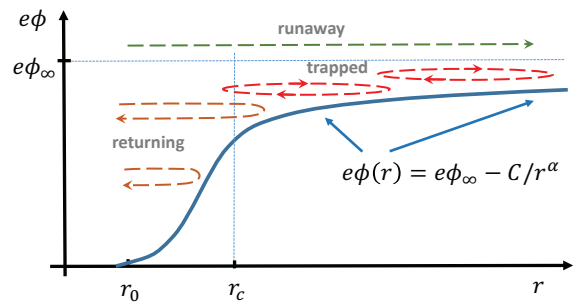}}	
    \caption{Sketch of the ambipolar electric potential energy outside the source $r\geq r_0$. The plot also shows the trajectories of some returning, trapped, and runaway electrons.}
    \label{potential}
\end{figure}

In order to understand the plasma dynamics outside the source region $r>r_0$, we need to describe the variation of the electric potential, $\phi(r)$, with the distance. In this respect, an important observation was made in \cite{konkashbaev1978,mirnov1979,ryutov2005} that most of the potential drop occurs {\em near} the plasma source. Indeed, {\em in a steady state}, the electrons leaving and returning back to the source by the potential barrier, should be in thermal equilibrium with the source. They should have the Maxwellian distribution with the same temperature as the source, and their density should decline with the distance according to the Boltzmann law $n_e(r)=n_0\exp\left(-e\phi(r)/T_{e, 0}\right)$. The ions, on the contrary, are not trapped by the potential but rather accelerated by it. However, their velocity does not change considerably by the potential difference, therefore, one can assume that they are streaming radially with, approximately, their thermal velocity $v_{T_{i, 0}}$, which is independent of the distance. Their density, therefore, declines as $n_i(r)=n_0(r_0/r)^2$, and from the condition of quasineutrality $n_i(r)=n_e(r)$ and Eq.~(\ref{eq1}), one obtains \cite[e.g.,][]{ryutov2005} that the potential nearly reaches its asymptotic value $\phi_\infty$ already at
\begin{eqnarray}
\label{r_crit}
r_c\sim r_0\left(\frac{m_i}{m_e}\right)^{1/4}.
\end{eqnarray}
For instance, in the case of the solar wind, we are dealing with a mostly hydrogen plasma, so this distances is $r_c \sim (m_i/m_e)^{1/4} r_0\approx  0.15 - 0.3\,\,{\rm AU}$. We may therefore predict that at $r\lesssim r_c$, the solar-wind electric field should be relatively strong compared to the electric field at large distances~$r \gg r_c$. The solid line in Figure~\ref{potential} sketches the electron potential energy as a function of the distance from the source. 
      
Beyond the point given by Eq.~(\ref{r_crit}), the potential cannot significantly change anymore. What happens to the electron distribution function after that point? This question was discussed in \cite[][]{konkashbaev1978}, see also \cite[][]{mirnov1979,ryutov2005}. It was realized that in the presence of very weak collisions, a significant fraction of the electrons at distances exceeding 
(\ref{r_crit}) would be accumulated, {in a steady state}, in the so-called {trapped} population, see Fig.~(\ref{potential}). 

In order to understand the origin of the trapped electron population,  we first note that in the absence of collisions, the electrons whose kinetic energies  exceed $e\phi(r)$, will not yet be turned back by the potential at the distance~$r$. A part of these electrons, whose energy is less than $e\phi_\infty$, eventually will be turned back to the source at larger distances; we will call these electrons the returning electrons.  The part whose energy is larger than $e\phi_\infty$ will stream to infinity unobstructed; we will call these electrons the runaway electrons. 

Now, what happens when weak Coulomb collisions are turned on? They will cause pitch angle scattering that can slightly increase the magnetic-field-perpendicular velocity of the streaming electrons. These electrons will not be able to return back to the source; they will be trapped by the increasing magnetic field at smaller distances and by the increasing electric potential at larger distances. Further pitch angle scattering will trap them even stronger. The pitch-angle diffusion will eventually fill up the ``reservoir" of trapped particles. In a steady state, it will be balanced by slow collisional losses through the boundary $E=e\phi_\infty$ due to diffusion in energy.  At any given distance $r\gg r_c$ the trapped electrons will tend to form a quasi-isotropic distribution. These electrons form the so-called ``core'' of the electron distribution function observed in the solar wind. 

The typical kinetic energy (the temperature) of this distribution can be estimated if one notices that the kinetic energy of the trapped population at a given distance $r$ cannot exceed $e\phi_\infty -e\phi(r)$, otherwise, such particles will stream to infinity. The boundary condition for the electron distribution function at $E=e\phi_\infty$ is, therefore, $f\equiv 0$. The temperature of the trapped electrons can, therefore, be estimated as $T_e(r)\sim e\phi_\infty - e\phi(r)$. If we assume (as can be checked later) that the electric potential at large distances has a power-law decline  (see Figure~\ref{potential}), the temperature of the trapped electrons should decline according to the same law, $T_e(r)\sim e\phi_\infty - e\phi(r)\propto 1/r^{\alpha}$. At the boundary of the narrow domain in the velocity space, $\sin^2\theta=v_\perp^2/v^2 \ll 1$, which separates the streaming electrons from the trapped ones, the isotropic distribution of the trapped electrons has to match the distribution of the non-trapped streaming electrons. In order to find the velocity distribution of the trapped electrons at some distance $r$, it is, therefore, necessary to know the velocity distribution of the streaming (``strahl'') electrons at the same distance. \\

\section*{The strahl electrons}
\label{strahl}
Let us assume that the plasma propagates along a magnetic flux tube  diverging radially from the source. The magnetic field strength and the ion density then decline with the distance as $B(r)=B_0(r_0/r)^2$ and $n_i(r)=n_0(r_0/r)^2$ (if necessary, these results can be generalized for an arbitrary expansion law $B(r)\propto n_i(r)\propto 1/r^{\gamma}$). The electron velocity distribution function $f(r,\mu, v)$ can then be described by the stationary drift-kinetic equation \cite[e.g.,][]{kulsrud2005}:
\begin{eqnarray}
\label{drift_kinetic}
\mu v \frac{\partial f}{\partial r} - \frac{1}{2}\frac{d\ln B}{dr}v \left(1-\mu^2\right) \frac{\partial f}{\partial \mu} -
\nonumber \\
 - \frac{e E_\parallel}{m_e} \left[\frac{1-\mu^2}{v} \frac{\partial f}{\partial \mu}  + \mu \frac{\partial f}{\partial v}\right] 
 = \hat C(f),
\end{eqnarray}
where $r$ is the distance along the magnetic field line, 
$\mu={\hat {\bf b}}\cdot {\bf v}/v=\cos\theta$ is the cosine of the angle between the electron velocity and the direction of the magnetic field line, $E_\|=-\nabla_\| \phi$ is the electric field along the magnetic field lines, and ${\hat C}(f)$ is the collision integral.   

\begin{figure}
\centerline{\includegraphics[width=\columnwidth]{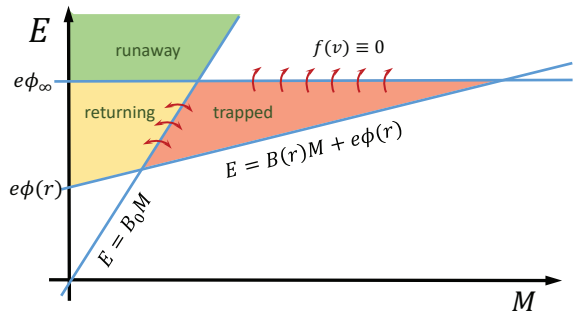}}	
    \caption{Energy - Magnetic moment ($E-M$) diagram for the electrons at distance $r$ from the source. The region defined by $E>B_0M$, $E>B(r)M+e\phi(r)$, and $E>e\phi_\infty$ is occupied by the electrons propagating from the source without collisions and escaping the ambipolar potential, the so-called {\em runaway} electrons. The region $E>B_0M$, $E>B(r)M+e\phi(r)$, and $E<e\phi_\infty$ corresponds to the free-streaming electrons that will be eventually turned back by the electric potential at distances larger than $r$ and stream back to the source, the so-called {\em returning} electrons. The region defined by $E<B_0M$, $E>B(r)M+e\phi(r)$, and $E<e\phi_\infty$, is not accessible by the free-streaming electrons. It can be only populated by the electrons scattered from the region $E>B_0M$ due to weak collisions (due to collisions, the electrons can also slowly escape the trapping potential and stream to infinity). These processes are indicated by the red arrows. The electrons in this region are {\em trapped}, they cannot stream back to the source due to the magnetic mirror and cannot run away due to the electric potential. In a steady state, they tend to form a quasi-spherical core of the electron distribution function.}
    \label{em_diagram}
\end{figure}

\begin{figure}
\centerline{\includegraphics[width=1.2\columnwidth]{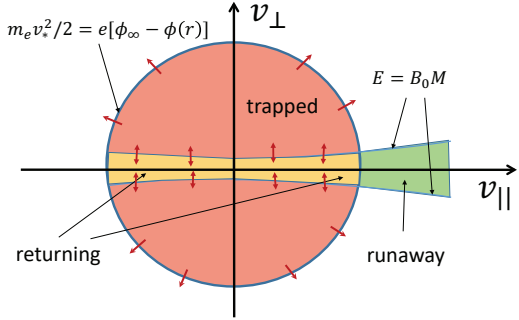}}	
    \caption{The same regions as in Figure~\ref{em_diagram} but plotted using the $v_\|-v_\perp$ variables. The boundaries separating the runaway, returning, and trapped electrons are defined by the same equations as in Figure~\ref{em_diagram}. The arrows show particle transport due to weak collisions. }
    \label{vv_diagram}
\end{figure}

Without collisions, $\hat C(f)=0$, the solution of the kinetic equation is an arbitrary function of the invariants of motion, the energy and the magnetic moment,  $E=m_e v^2/2+e\phi(r)$ and $M=m_e v^2(1-\mu^2)/(2B(r))$. {Indeed, written in the new variables $E$, $M$, and $r$, the drift-kinetic equation (\ref{drift_kinetic}) has a remarkably simple form $\mu v\partial f(E, M, r)/\partial r={\hat C}(f)$ (where $\mu$ and $v$ have to be expressend through $E$ and $M$ as well). The collisionless equation then reads $\partial f(E, M, r)/\partial r=0$, which means that the distribution function is independent of the distance, $f(E, M, r)=f(E, M)$.} Since at the source this function matches the Maxwellian, we get for the distribution function of streaming particles ($v_\|>0$) at a distance $r>r_0$ from the source~\cite[e. g.,][]{boldyrev19}:
\begin{eqnarray}\label{f_beam}
& f(r,E,M)=A_0\exp\left\{-\frac{E}{T_{e, 0}} \right\}\theta\left(E-M B_0\right) \nonumber \\
& = A_0\exp\left\{-\frac{e\phi(r)}{T_{e, 0}} -\frac{m_ev^2}{2T_{e, 0}}\right\}\theta\left(v^2+\frac{2e\phi(r)}{m_e}-\frac{B_0}{B(r)}v_\perp^2 \right),
\end{eqnarray} 
where the normalization coefficient is $A_0=n_0(m_e/2\pi T_{e, 0})^{3/2}$, and $\theta$ is the Heaviside step function. At the distance $r>r_0$, the regions of returning, trapped, and runaway electrons look in the new variables as shown in Figure~\ref{em_diagram}. For convenience, Figure~\ref{vv_diagram} shows the same regions in the original $v_\|$ and $v_\perp$ coordinates.

From the $\theta$-function in Eq.~(\ref{f_beam}), we see that when $B(r)\ll B_0$, the streaming particles occupy a narrow domain in the phase space defined by
\begin{eqnarray}
v_\perp^2 <  \frac{B(r)}{B_0}\left(v^2+\frac{2e}{m_e}\phi(r)\right)\approx \frac{B(r)}{B_0}\left(v_\|^2+\frac{2e}{m_e}\phi(r)\right).\label{cone}
\end{eqnarray} 
If the decline in the magnetic field strength is sufficiently large,  Eq.~(\ref{cone}) corresponds to a very narrow cone in the velocity space.  Indeed, let us estimate the width of this cone for the particles that will be turned back by the potential. For such particles, $m_e v^2/2\lesssim e\phi_\infty -e\phi(r) \sim T_e(r)$, while from Eq.~(\ref{cone}) we estimate $m_ev_\perp^2/2\sim (B(r)/B_0)e\phi_\infty$. We, therefore, obtain
\begin{eqnarray}\label{cone_0}
\sin^2\theta \equiv v_\perp^2/v^2\sim \frac{B(r)}{B_0}\frac{e\phi_\infty}{\left(e\phi_\infty-e\phi(r)\right)}\sim \frac{B(r)}{B_0}\frac{e\phi_\infty}{T_e(r)}\ll 1.
\end{eqnarray}
 
If the collisions are completely neglected, the region of the phase space outside the cone defined by Eq.~(\ref{cone_0}), cannot be accessed by the electrons traveling from the source. This region can, however, be populated by the electrons {\em scattered} from the cone~(\ref{cone_0}) to larger $v_\perp$ by weak Coulomb collisions, see Fig.~(\ref{em_diagram}). If the energy of the scattered electrons is such that $m_ev^2/2< e\phi_\infty-e\phi(r)$, these electrons will form the  trapped population. They will not be able to stream back to the source due to the magnetic mirror at small distances, and they will not be able to leave the system due to the potential barrier at large distances. 

The trapped electrons tend to form an isotropic distribution. At the energies exceeding the trapping potential, $m_ev^2/2 \geq m_e v_*^2/2=e(\phi_\infty-\phi(r))$, there are no trapped electrons, and their  distribution function should satisfy the boundary condition $f_{tr}=0$. {As discussed in \cite{konkashbaev1978,ryutov2005}, at the boundary of the cone~(\ref{cone_0}), the trapped-electron distribution function, $f_{tr}$, should, in addition,  be matched with the streaming-electron function~(\ref{f_beam}). The level of the trapped population is therefore a result of  two competing processes - particle influx from the streaming population due to pitch-angle scattering and particle losses through the boundary $v=v_*$ due to energy diffusion. For Coulomb collisions, the rate of electron pitch-angle scattering is equal to the rate of energy scattering due to electron-electron collisions~\cite[e.g.,][]{nrl}. Because of this fact, it is  reasonable to expect that the level of the trapped population will be comparable to that of the streaming population. (This result is, however, not immediately obvious. Its justification and a more direct derivation is given in the Appendix.) One can therefore approximate the trapped function as:
\begin{eqnarray}\label{f_tr}
f_{tr}(x, {\bf v})\approx A_0\exp\left\{-\frac{e\phi_\infty}{T_{e, 0}} \right\}\theta(v_*-v)\nonumber \\
\approx A_0\left(\frac{m_e}{m_i}\right)^{1/2}\theta(v_*-v).
\end{eqnarray}
In this equation, we substituted $E\approx e\phi_\infty$ and used an approximate relation following from Eq.~(\ref{eq1}):
\begin{eqnarray}
\exp\left\{-\frac{e\phi_\infty}{T_{e, 0}} \right\}\approx \left(\frac{m_e}{m_i} \right)^{1/2}\frac{\left(T_{e, 0}T_{i, 0}\right)^{1/2}}{e\phi_\infty}\approx \left(\frac{m_e}{m_i} \right)^{1/2}.
\end{eqnarray}
We also omitted numerical factors of order unity. 
{Formula~(\ref{f_tr}) should be understood as a model, since the physical distribution function cannot have a sharp gradient in the velocity space. Rather, due to the electron-electron collisions, the electrons will diffuse through the boundary $v=v_*$ and the gradient of the distribution function will be smoothed out over a scale comparable to~$v_*$. The simple model expression~(\ref{f_tr}) will however suffice for the dimensional estimates used below.} 

With the aid of Eq.~(\ref{f_tr}) one can find the density of the trapped electrons as $n_{tr}=\int f_{tr} d^3\, v$, and equating it to the ion density at the same position, $n(r)=n_0(r_0/r)^2$, one can find the typical velocity of the electron distribution,~$v_*$. The ``temperature'' of the trapped electrons is then \cite[][]{mirnov1979,ryutov2005}:
\begin{eqnarray}\label{t_collisionless}
T_e(r)\sim  \frac{m_e v_*^2}{2} \sim T_{e, 0} \left(\frac{m_i}{m_e} \right)^{1/3} \left(\frac{r_0}{r}\right)^{4/3},
\end{eqnarray}
where $T_{e, 0}$ is the electron temperature at the source.  
The predicted temperature drop with the distance, $T_e(r)\propto r^{-4/3}$, is relatively steep. In particular, it is steeper than the observational temperature profiles in the solar wind. 

In the next section we demonstrate that given a large enough length of the expansion region, the solution given by Eq.~(\ref{t_collisionless}) will inevitably break down beyond a certain scale, no matter how weak the collisions are. The system will then approach a new asymptotic state, where the narrow collimation angle is regularized by collisions, and where the temperature profile is significantly less steep. The interval of scales before the break point will be called the {\em near zone}. This is the interval where the solution described by Eqs.~(\ref{f_tr}, \ref{t_collisionless}) is applicable. The interval of scale beyond this point will be called the {\em far zone}. It is discussed in the next section.\\

\section*{Solution in the far zone}
\label{farzone}
In the preceding section, the collisions were assumed to be too weak to alter the shape of the free-streaming electron beam given by Eq.~(\ref{f_beam}). However, if the travel distance of the electrons becomes large enough, the collimation angle   described by Eq.~(\ref{cone}) becomes extremely narrow so that the collisional broadening of the beam cannot be neglected. In order to study this process, we need to take into account Coulomb collisions in the drift-kinetic equation~(\ref{drift_kinetic}). 

The main collisional process we are interested in is pitch-angle scattering. To study this process, we retain in the collision integral only the pitch-angle scattering part. We assume that the scattering is given by collisions with the ions and with the electrons forming nearly isotropic trapped populations (the beam population is rather tenuous and it does not significantly contribute to the electron density). 
For the strahl electrons whose energy is comparable to the thermal energy, the pitch-angle scattering operator can be estimates by order of magnitude as \cite[e.g.,][]{helandersigmar02}:
\begin{eqnarray}\label{coll_op_eq_th}
\hat C(f)  \sim \left(\frac{4 \pi n(r) e^4 \Lambda  }{m_e^2 v^3_{th, e}(r)}\right) \frac{\partial}{\partial \mu} \left(1-\mu^2\right)\frac{\partial f}{\partial \mu}.
\end{eqnarray}
In this equation, $\Lambda$ denotes the standard Coulomb  logarithm,  $n(r) = n_e(r) =n_i(r)$ is the density of the ions and the trapped electrons.

Let us rewrite the drift-kinetic equation~(\ref{drift_kinetic}) for the particles streaming from the sun ($v_\|>0$), using the new variables
 $r$, $E$, and $M$ \cite[see, e.g.,][]{boldyrev19}:
\begin{eqnarray}
\frac{\partial f(r, E, M)}{\partial r} =\frac{4 \pi e^4 \Lambda   n(r)}{{\cal E}(E,r)B(r)}\frac{\partial }{\partial M}M\sqrt{1-\frac{MB(r)}{{\cal E}(E,r)}} \frac{\partial f}{\partial M},\label{diffusion}
\end{eqnarray}
where ${\cal E} (E, r)=E-e\phi(r)=m_ev^2/2$. It is easy to see that the ratio appearing in the square root of Eq.~(\ref{diffusion}) is related to the propagation angle, ${MB(r)}/{{\cal E}(E,r)}=v_\perp^2/v^2=\sin^2\theta$. We assume that this angle is small and neglect this term in our estimates. 

Let us consider the electrons streaming from the source, that will eventually be turned back by the potential barrier. These electrons can become trapped after pitch angle scattering. For these electrons we have $m_ev^2/2< e(\phi_\infty - \phi(r))$, in which case we may estimate their total energy at a distance $r\gg r_c$ as $E\approx e\phi_\infty$. Analogously to the method also applied in \cite{horaites2019,boldyrev19}, we can introduce a new variable according to
\begin{eqnarray}
dy=\left(\frac{4\pi e^4\Lambda n(r) }{{\cal E}(r) B(r)} \right)\,dr.\label{dy}
\end{eqnarray} 
In the case of plasma outflow along a radially expanding magnetic flux tube, $B(r)/n(r)={\rm const}$. For definiteness, we may express this constant through the magnetic-field and density values at $r=r_0$. By integrating Eq.~(\ref{dy}), we then get
\begin{eqnarray}
y=\left(\frac{4\pi e^4\Lambda   n_0}{(1+\alpha)B_0{\cal E}(r)} \right) \, r,
\end{eqnarray}
where we took into account that at $r\gg r_c$, we can approximate ${\cal E}(r)\approx C/r^{\alpha}$. The simplified equation~(\ref{diffusion}) now takes the form:
\begin{eqnarray}
\frac{\partial f}{\partial y} =\frac{\partial }{\partial M}M \frac{\partial f}{\partial M},\label{diffusion_simple}
\end{eqnarray}
which is equivalent to the radial diffusion equation in a 2D plane; this can be verified by the change of variable $\zeta=\sqrt{M}$.

The diffusion equation~(\ref{diffusion_simple}) can be solved if we specify the distribution function at the initial position~$y_c$, which can be chosen to correspond to the distance $r_c$ from the source.  The distribution function at the point $r_c$ is given by formula~(\ref{f_beam}), which we can conveniently rewrite as \cite[see e.g.,][]{boldyrev19}: 
\begin{eqnarray}
f(r_c, E, M)=A_0\exp\left\{-\frac{E}{T_{e,0}} \right\}\frac{E}{B_0}\Delta(M),
\end{eqnarray}
where $\Delta(M)$ is a localized function of $M$, whose width is $\delta M_0\approx {E}/{B_0}$ and amplitude $B_0/E$. This function, therefore, tends to a delta function when $E/B_0\to 0$. At larger distances $r\gg r_c$, this function will broaden due to the diffusion.  If the diffusion broadening is much larger than the initial width, that is, if $y\gg E/B_0$,  the function $\Delta(M)$ will be replaced by the solution of the diffusion equation, 
\begin{eqnarray}
\Delta(M)\to \frac{1}{y}\exp\left\{-\frac{M}{y} \right\}. 
\end{eqnarray}

In the previous treatments of the problem \cite[][]{konkashbaev1978,mirnov1979,ryutov2005}, this diffusion broadening was neglected. Let us estimate when this can be done, which will allow us to establish the limits of validity
 of the near-zone solution. The diffusion broadening, $\delta M_y\sim y$, is negligible when $\delta M_y \ll \delta M_0$, which can be written as
\begin{eqnarray}\label{comparison}
\frac{\delta M_y}{\delta M_0}\sim \frac{4\pi e^4\Lambda   n(r) B_0 r}{B(r){\cal E}(r)E }\sim \left(\frac{{\cal E}(r)}{E}\right)\left(\frac{B_0}{B(r)}\right)\left(\frac{r}{\lambda_{\rm mfp}}\right) \nonumber \\
\sim \left(\frac{T_e(r)}{e\phi_\infty}\right)\left(\frac{B_0}{B(r)}\right)\left(\frac{r}{\lambda_{\rm mfp}}\right)\ll 1,
\end{eqnarray}
where we introduced the electron mean-free path as:
\begin{eqnarray}\label{e_mfp}
{\lambda_{\rm mfp}}=\frac{{T}_e^2(r) }{4\pi e^4\Lambda   n(r)}.
\end{eqnarray}
Recalling the definition of the collimation angle in the near zone, given by Eq.~(\ref{cone_0}), we can re-write the applicability criterion~(\ref{comparison}) for the near-zone solution as
\begin{eqnarray}\label{validity}
\frac{r}{\lambda_{\rm mfp}}\ll \sin^2{\theta} \ll 1.
\end{eqnarray}
This is the criterion of the near-zone approximation mentioned in the introduction section. We can see that the restriction on the mean-free path given by Eq.~(\ref{validity}) becomes less satisfied at large  distances, since as we have previously established for this case, $T_e(r)\propto r^{-4/3}$, while $B(r)\propto r^{-2}$. While this is not an issue for the laboratory mirror machines, where the electron mean-free path exceeds the length of the expansion region by about five orders of magnitude, this is not necessarily a situation in natural plasmas. 

It is, therefore, important to consider a less restrictive case
\begin{eqnarray}\label{validity_new}
\left(\frac{B(r)}{B_0}\right)\left(\frac{e\phi_\infty}{T_e(r)}\right)\lesssim \left(\frac{r}{\lambda_{\rm mfp}}\right)\ll 1,
\end{eqnarray}
when the electron-beam width is dominated by the diffusion rather than the free-streaming collimation effect. As a result, we obtain:
\begin{eqnarray}\label{f_total}
f=A_0\exp\left\{-\frac{E}{T_{e,0}}\right\}\frac{E}{B_0}\times \frac{1}{y}\exp\left\{-\frac{M}{y}\right\}.
\end{eqnarray}
This expression can be simplified if we recall that the particles reaching  very large distances $r\gg r_c$ before being reflected by the potential barrier, have $E\approx e\phi_\infty$, and we can use Eq.~(\ref{eq1}) to replace, approximately,
\begin{eqnarray}
E\exp\left\{-\frac{E}{T_{e,0}}\right\}\approx T_{0,e}\left(\frac{m_e T_{i, 0}}{m_i T_{e, 0}} \right)^{1/2}.
\end{eqnarray}
In addition, we can approximate ${\cal E}(r)\approx T_e(r)$, so that expression~(\ref{f_total}) is, approximately, 
\begin{eqnarray}
\label{f_total_approx}
f\approx A_0\frac{T_e(r)T_{0,e}}{4\pi e^4\Lambda  n_0\, r}\left(\frac{m_e T_{i, 0}}{m_i T_{e, 0}}\right)^{1/2}\exp\left\{-\frac{M}{y}\right\}.
\end{eqnarray}
This is the distribution function of the electrons streaming from the sun ($v_\|>0$), that have not yet been turned back by the potential barrier. 

{In order to find the distribution of the trapped electrons, we have to find a solution for which the inflow of the pitch-angle-scattered streaming particles balances the diffusive losses at the trapping boundary~$v= v_*$. Analogously to the derivation of Eq.~(\ref{f_tr}), it is reasonable to expect that the level of the trapped population is comparable to that of the streaming population~(\ref{f_total_approx}), and to model the isotropic distribution function of trapped particles as:
\begin{eqnarray}\label{f_tr_approx}
f_{tr}\approx A_0\frac{T_e(r)T_{e,0}}{4\pi e^4\Lambda  n_0\, r}\left(\frac{m_e T_{i, 0}}{m_i T_{e, 0}}\right)^{1/2}\theta(v_* -v).
\end{eqnarray}
}
{A more straightforward derivation of this result can be found in the Appendix.} 

Using expression~(\ref{f_tr_approx}) we can now calculate the density of the trapped electrons and equate it to the density of the ions, which gives
\begin{eqnarray}
\left(\frac{m_e}{2\pi T_{e, 0}}\right)^{3/2}\frac{T_e(r)T_{0,e}}{4\pi e^4\Lambda  n_0\, r}\left(\frac{m_e T_{i, 0}}{m_i T_{e, 0}}\right)^{1/2}\frac{4\pi}{3}v_*^3=\left(\frac{r_0}{r}\right)^2.
\end{eqnarray}
From that we can find the temperature of the trapped electrons
\begin{eqnarray}\label{T_collisional}
T_e(r)\sim 
T_{e, 0}\left(\frac{r_0}{r} \right)^{2/5}\left(\frac{m_i}{m_e}\right)^{1/5}\left[\frac{ 3 \pi^{3/2} e^4\Lambda   n_0 r_0 }{ T_{i, 0}^2} \right]^{2/5}.
\end{eqnarray}
The expression in the square brackets is just $(T_{e, 0}/T_{i, 0})^2 (r_0/\lambda_{{\rm mfp}, 0})$, which is a parameter defined by the plasma conditions at the source. In the case of mirror devices, this parameter is significantly smaller than one, while in the case of the solar wind it happens to be on the order of $0.1$ -- $1$.  As a result we estimate
\begin{eqnarray}\label{t_collisional}
T_e(r)\sim T_{e, 0}\left(\frac{m_i}{m_e}\right)^{1/5}\left(\frac{T_{e, 0}}{T_{i, 0}}\right)^{4/5}\left(\frac{r_0}{\lambda_{{\rm mfp}, 0}} \right)^{2/5}\left(\frac{r_0}{r} \right)^{2/5}.
\end{eqnarray}
An important difference of this result from the collisionless expression~(\ref{t_collisionless})  is the  scaling of the temperature with the distance, $T_e(r)\propto r^{-2/5}$ (so that $\alpha=2/5$). Note that it is in a good agreement with the slower than adiabatic radial temperature decline in the solar wind.

Finally, we need to verify that the assumptions used in our derivation are satisfied by our solution. Our derivation essentially assumed that the collimation angle of the electron beam is small. The width of the collimation angle can be estimated from  $M\sim y$, which gives
\begin{eqnarray}\label{angle}
\sin^2\theta\sim \frac{yB(r)}{{\cal E}(r)}\approx \frac{4\pi e^4\Lambda   n_0 B(r) \, r}{B_0{T_e}^2(r)}\approx \frac{r}{\lambda_{\rm mfp}}.
\end{eqnarray}
In order for our theory to be applicable, we need to require that the collimation angle~(\ref{angle}) is much smaller than one. We therefore arrive at the applicability criterion for the far-zone solution 
\begin{eqnarray}\label{validity_new_ref}
\sin^2\theta\sim \frac{r}{\lambda_{\rm mfp}}\ll 1,
\end{eqnarray}
which we have mentioned in the introduction. It is important to note that due to the obtained scaling of the temperature~(\ref{t_collisional}) and the scaling of the magnetic field $B(r)\propto r^{-2}$, the collimation angle in Eq.~(\ref{angle}) {\em decreases} with the distance. Therefore, in contrast with the collisionless case~(\ref{validity}), the applicability criterion of the far-zone solution expressed by Eqs.~(\ref{validity_new}) or~(\ref{validity_new_ref}), does not break down but rather becomes even better satisfied at larger distances. 

The transition scale between the near zone and the far zone, $r_*$, can be established from comparison of Eq.~(\ref{t_collisionless}) and Eq.~(\ref{t_collisional}),
\begin{eqnarray}
\frac{r_*}{r_0}\sim \left(\frac{m_i}{m_e}\right)^{1/7}\left(\frac{T_{0,i}}{T_{0,e}} \right)^{6/7}\left(\frac{\lambda_{{\rm mfp}, 0}}{r_0} \right)^{6/7}.
\end{eqnarray}
For instance, in the solar wind, $r_*$ is just a few times larger than~$r_0$, so it is the far zone solution that is relevant for the solar wind.\\

\section*{Conclusions}
We have proposed a theory that relates, in a steady state, the electron distribution function of the hot thermal source with the distribution function in the nearly collisionless expansion region outside of the source. The theory has several important predictions that, we argue, are relevant for the electron-temperature distribution in the inner heliosphere. 

First, it predicts that the temperature of the electrons in the collisional thermal source (the solar corona in our application) should be smaller than the temperature of the ions.  Even in the absence of preferential ion heating mechanisms often invoked in theories of coronal heating \cite[e.g.,][]{chandran10,klein2016}, the theory predicts that the electrons should be about $3.5$~times colder than the ions, a result quite close to the observations \cite[e.g.,][]{mercier2015}. This result is a natural consequence of the fact that in a steady state, the electrons streaming out of the source must carry the same energy flux as the ions~\cite[][]{konkashbaev1978,ryutov2005}. We thus explain the observed electron and ion temperature difference in the solar corona not by preferential heating of the ions, but rather by preferential cooling of the electrons. 

Second, our theory naturally predicts the formation of a highly collimated electron strahl and a nearly isotropic  electron core. The core in our model is the population of the so-called trapped electrons, which bounce in a well  formed by the electric potential at larger distances and the magnetic mirror at smaller, and, therefore, have a chance to isotropize by collisions. The core consists of the electrons that have been stripped from the strahl by weak collisions. The typical energy (temperature) of the core is predicted in our theory to decline as a power law with the heliospheric distance, $T_e(r)\propto r^{-2/5}$. This behavior is rather close to the observational temperature profile of the fast solar wind in the inner heliosphere, and it is broadly consistent with the shallower-than-adiabatic temperature profile of the slow wind. Our explanation of the solar wind temperature is related to the solar wind heating by the electron flux; we argue that the energy carried by the streaming electrons is sufficient to explain the observational non-adiabatic temperature profile. Our explanation is complementary to the modes invoking plasma heating by turbulence \cite[e.g.,][]{stawarz2009,cranmer2014}.   

Finally, in our model, the magnetic field configuration is assumed to be stationary, since the thermal velocity of the electrons is higher than the speed of the solar wind. The population of the trapped electrons is also stationary in our model, that is, it does not propagate away from the source. In the solar wind, however, the electron core population is not stationary but rather moving away from the sun with the speed comparable to that of the solar wind. Since our theory neglects the solar wind speed as compared to the thermal speed of the electrons, the bulk electron motion is, therefore, a higher-order effect with respect to the small parameter $v_{sw}/v_{th, e}\ll 1$, and it does not change our results. 

It is, however, interesting to discuss what causes the electron-core drift. A possible explanation, in our view, is reflected in the fact that the ion-electron distribution where the electron and ion bulk velocities are significantly shifted with respect to each other, is unstable. As was recently shown in \cite{horaites18b} (see also earlier studies in \cite[][]{perkins1973}), when this shift is on the order of the Alfv\'en velocity, it generates Alfv\'en waves with $k\sim 1/d_i$. We speculate that the interaction of the electrons with the resulting turbulent fluctuations  drags the electrons together with the ions, as to minimize the velocity mismatch between the two species, and to reduce the instability. 

More quantitatively, consider a trapped electron bouncing between a potential wall (at some large distance, $r_2$) and a magnetic mirror (at some small distance, $r_1$). Let us assume for simplicity that due to an interaction with a wave the electron parallel velocity $v_\|$  increases slightly, while the perpendicular velocity remains the same.  Since the electron energy increases, the electron will be turned back by the potential at a slightly larger radial distance; in particular, the turning point will shift by $\Delta r_2/r_2\approx (5/2)\Delta E/T_e(r)$ away from the sun. The mirror reflection point, on the contrary, will shift toward to the sun by $\Delta r_1/r_1\approx (1/2)\Delta E/E$~\footnote{In these estimates, we have used the fact that the mirror reflection point is defined by the equation 
$E=e\phi(r_1)+MB(r_1)\approx MB(r_1)$, while the potential reflection point by the equation $E\approx e\phi(r_2)\approx e\phi_\infty -T_e(r_2)$.  Taking into account that $T_e(r)\propto r^{-2/5}$ and $B(r)\propto r^{-2}$, we estimate that, by absolute value, $\Delta r_1/r_1\approx (1/2)\Delta E/E$, while $\Delta r_2/r_2\approx (5/2)\Delta E/T_e(r_2)$.}. Since $r_2\gg r_1$ and $E\gg T_e(r_2)$, we see that the mirror-reflection-point shift is negligible, $\Delta r_1\ll \Delta r_2$, and the net shift of the electron trajectory is {\em away} from the sun. Therefore, due to energization by resonant collisions with waves, the trapped electrons will drift away from the sun. This mechanism explains the drift of the electron core together with the ions, and it also explains why the established mismatch between the two bulk velocities is comparable to the Alfv\'en speed. \\

\appendix
\section*{Appendix: The trapped electrons}
\label{appendix}
{In this section, we give a more straightforward kinetic derivation of the trapped electron population, and in particular, justify the estimates given in Eqs.~(\ref{f_tr}) and~(\ref{f_tr_approx}). Our derivation is based on the following  observations. First, we integrate the steady-state drift-kinetic equation,
\begin{eqnarray}
\left(v_\|\frac{\partial f}{\partial r}\right)_{E,M}={\hat C}(f),
\end{eqnarray}
over the trapped region in the velocity space, $v<v_*$. We then notice that the term
\begin{eqnarray}\label{source}
\int\limits^{v_*}_0 \left(v_\|\frac{\partial f}{\partial r}\right)_{E,M}d^3 v\equiv \nonumber \\
\equiv \int\limits^{e\phi_\infty}_{e\phi(r)}dE \int\limits^{\frac{E-e\phi(r)}{B(r)}}_{0} dM\,\frac{2\pi B(r)}{m_e^2\left|v_\| \right| }\left(v_\|\frac{\partial f}{\partial r}\right)_{E,M}  
\end{eqnarray} 
has the meaning of the number of particles supplied to the trapped population per unit time at a given position~$r$. We call this term the ``source integral.'' 

Then, the distribution function in the region $v<v_*$ may be modeled as the sum of two components: the population of particles traveling from the sun plus the population of trapped electrons. Obviously, the trapped particles and the reflected particles traveling back to the sun, do not contribute to the integral in Eq.~(\ref{source}) as their distribution functions are symmetric in~$v_\|$. The contribution to this integral will, therefore, be given by the particles propagating directly from the sun that will become trapped after the first reflection at the potential barrier.} We evaluate this integral separately for the particles in the near and far zones. 

In the near zone, the distribution of particles propagating from the sun is given according to Eq.~(\ref{f_beam}) and Eq.~(\ref{diffusion_simple}), by the function
\begin{eqnarray}\label{stream_near}
f(r,E,M) =A_0\exp\left\{-\frac{E}{T_{e, 0}} \right\}{\tilde \theta}\left(E/B_0-M\right),
\end{eqnarray}
where ${\tilde \theta}$ denotes the Heaviside theta-function whose sharp boundary has been smoothed by pitch-angle diffusion over a narrow layer of width~$\delta M\approx y$. (We remind the reader that in the near zone, the pitch-angle diffusion effects do not considerably distort the distribution function of streaming electrons.) Obviously, only the streaming particles within this layer will become trapped after first reflection, while the particles with $M<E/B_0$ will travel all the way back to the sun. The source integral in Eq.~(\ref{source}) can, therefore,  be evaluated as 
\begin{eqnarray}\label{source_near}
\int\limits^{v_*}_0 \left(v_\|\frac{\partial f}{\partial r}\right)_{E,M}d^3 v\approx \nonumber \\
\approx \int\limits^{e\phi_\infty}_{e\phi(r)}dE \int\limits^{\frac{E}{B_0}+\delta M}_{\frac{E}{B_0}} dM\,\frac{2\pi B(r)}{m_e^2\left|v_\| \right| }\left(v_\|\frac{\partial f}{\partial r}\right)_{E,M}.  
\end{eqnarray}
Substituting here the function given by Eq.~(\ref{stream_near}), we evaluate the source integral in the near zone as:
\begin{eqnarray}\label{source_near_final}
\int\limits^{v_*}_0 \left(v_\|\frac{\partial f}{\partial r}\right)_{E,M}d^3 v\approx -A_0\frac{8\pi^2 n_0 e^4 \Lambda B(r)}{m_e^2 B_0}\left(\frac{m_e}{m_i}\right)^{1/2}.
\end{eqnarray}

In the far zone, nearly {\em all} the streaming electrons described by Eq.~(\ref{f_total_approx}) will become trapped already after the first reflection by the potential barrier. Indeed, their angular broadening is significantly larger than that given by the collisionless formula~(\ref{cone}). The distribution function in the region $v<v_*$ is, therefore, the sum of the population of the first-time passing particles given by Eq.~(\ref{f_total_approx}) plus the population of trapped electrons. The trapped particles do not contribute to the integral in Eq.~(\ref{source}) as their distribution is symmetric with respect to ~$v_\|$, so the contribution to this integral comes solely from the {streaming } population. Substituting Eq.~(\ref{f_total_approx}) into Eq.~(\ref{source}) and evaluating the integral, we thus obtain the source integral  for the far zone:
\begin{eqnarray}\label{source_far_final}
\int\limits^{v_*}_0 \left(v_\|\frac{\partial f}{\partial r}\right)_{E,M}d^3 v\approx -A_0\frac{2\pi T_e(r)T_{e,0}B(r)}{m_e^2B_0 r}\left(\frac{m_eT_{i, 0}}{m_i T_{e, 0}}\right)^{1/2}.
\end{eqnarray}

{The next step is to note that in a steady state, the source integrals given by Eq.~(\ref{source_near_final}) or Eq.~(\ref{source_far_final}) have to be balanced by the integral of the collisional term, 
\begin{eqnarray}\label{balance}
\int\limits_0^{v_*}{\hat C}(f_{tr})\,d^3 v,
\end{eqnarray}
which describes particle loss through the boundary at $v=v_*$. In the collisional integral, the dominant contribution is provided by the {\em trapped} particles, since the number of particles
in the streaming population is significantly smaller.} {In the case of isotropic distribution of trapped electrons, the collisional integral has the form
\begin{eqnarray}
{\hat C}(f_{tr})=\frac{4\pi e^4 \Lambda}{m_e^2}\frac{1}{v^2}\frac{\partial }{\partial v}\left[D\frac{\partial f_{tr}}{\partial v}+Ff_{tr} \right],
\end{eqnarray}
where
\begin{eqnarray}
D=\int\limits_0^v\frac{v'^2}{3v}f_{tr}(v')\,d^3v'+\int\limits_v^\infty\frac{v^2}{3v'}f_{tr}(v')\,d^3v',
\end{eqnarray}
and
\begin{eqnarray}
F=\int\limits_0^vf_{tr}(v')\, d^3v'.
\end{eqnarray}
In this integral, only the electron-electron collisions are retained as the electron-ion collisions lead to a significantly weaker energy exchange between the particles. 
The integral in Eq.~(\ref{balance}) will therefore give:
\begin{eqnarray}
\int\limits_0^{v_*}{\hat C}(f_{tr})d^3 v=\frac{(4\pi)^2e^4\Lambda}{m_e^2}\left[D_*\frac{\partial f_{tr}}{\partial v_*}\right],
\end{eqnarray}
where we denoted 
\begin{eqnarray}
D_*=\int\limits_0^{v_*}\frac{v'^2}{3v_*}f_{tr}(v')\,d^3v'\equiv \frac{n(r)}{v_*} v_{th}^2\approx {n(r)v_{th}},
\end{eqnarray}
and took into account that $f_{tr}\equiv 0$ for $v\geq v_*$.} {Evaluating the derivative of the distribution function  as $\partial f_{tr}/\partial v_*\approx -f_{tr}/v_{th}$, we finally estimate
\begin{eqnarray}\label{C_integral}
\int\limits_0^{v_*}{\hat C}(f_{tr})d^3 v\approx -\frac{(4\pi)^2n(r)e^4\Lambda}{m_e^2}f_{tr}.
\end{eqnarray}

Comparing this result with the near-zone source integral, Eq.~(\ref{source_near_final}), we recover, up to a numerical factor, the estimate for the trapped distribution function $f_{tr}$ given in Eq.~(\ref{f_tr}). Similarly, comparing Eq.~(\ref{C_integral}) with the source integral in the far zone, Eq.~(\ref{source_far_final}), we obtain the estimate for the trapped function that coincides with the result of Eq.~(\ref{f_tr_approx})}.

The similarity of the results obtained by the two different derivations is not surprising. Indeed, for Coulomb collisions, the coefficient of pitch-angle electron diffusion used to derive Eq.~(\ref{f_total_approx}) and Eq.~(\ref{stream_near}) is equal to the coefficient of electron-electron energy diffusion used to derive Eq.~(\ref{C_integral}). Since in a steady state, the inflow of the pitch-angle scattered electrons into the trapped domain is equal to their loss from this domain through the potential boundary, the resulting levels of the streaming and trapped populations are proportional to each other. \\

\section*{Acknowledgements}
{SB is grateful to Konstantinos Horaites for the many discussions of the physics of the electron strahl in the solar wind, and to Vladimir Mirnov and Azamat Zhaksylykov for the discussions of the \citet{konkashbaev1978} theory at the early stages of this project.  The work of SB was supported by the NSF under the grant no. NSF PHY-1707272 and by NASA under the grant no. NASA 80NSSC18K0646. CBF, JE, and SB are also supported by the Wisconsin Plasma Physics Laboratory (DOE grant No. DE-SC0018266).}




\bibliographystyle{mnras}

\begin{thebibliography}{}
\makeatletter
\relax
\def\mn@urlcharsother{\let\do\@makeother \do\$\do\&\do\#\do\^\do\_\do\%\do\~}
\def\mn@doi{\begingroup\mn@urlcharsother \@ifnextchar [ {\mn@doi@}
  {\mn@doi@[]}}
\def\mn@doi@[#1]#2{\def\@tempa{#1}\ifx\@tempa\@empty \href
  {http://dx.doi.org/#2} {doi:#2}\else \href {http://dx.doi.org/#2} {#1}\fi
  \endgroup}
\def\mn@eprint#1#2{\mn@eprint@#1:#2::\@nil}
\def\mn@eprint@arXiv#1{\href {http://arxiv.org/abs/#1} {{\tt arXiv:#1}}}
\def\mn@eprint@dblp#1{\href {http://dblp.uni-trier.de/rec/bibtex/#1.xml}
  {dblp:#1}}
\def\mn@eprint@#1:#2:#3:#4\@nil{\def\@tempa {#1}\def\@tempb {#2}\def\@tempc
  {#3}\ifx \@tempc \@empty \let \@tempc \@tempb \let \@tempb \@tempa \fi \ifx
  \@tempb \@empty \def\@tempb {arXiv}\fi \@ifundefined
  {mn@eprint@\@tempb}{\@tempb:\@tempc}{\expandafter \expandafter \csname
  mn@eprint@\@tempb\endcsname \expandafter{\@tempc}}}

\bibitem[\protect\citeauthoryear{{Bale} et~al.,}{{Bale}
  et~al.}{2016}]{bale2016}
{Bale} S.~D.,  et~al., 2016, \mn@doi [Space Science Reviews]
  {10.1007/s11214-016-0244-5}, \href
  {http://adsabs.harvard.edu/abs/2016SSRv..204...49B} {204, 49}

\bibitem[\protect\citeauthoryear{{Ber{\v{c}}i{\v{c}}}, {Maksimovi{\'c}}, {},
  {Land i}  \& {Matteini}}{{Ber{\v{c}}i{\v{c}}} et~al.}{2019}]{bercic2019}
{Ber{\v{c}}i{\v{c}}} L.,  {Maksimovi{\'c}} {} M.,  {Land i} S.,   {Matteini}
  L.,  2019, \mn@doi [\mnras] {10.1093/mnras/stz1007}, \href
  {https://ui.adsabs.harvard.edu/abs/2019MNRAS.486.3404B} {486, 3404}

\bibitem[\protect\citeauthoryear{{Boldyrev} \& {Horaites}}{{Boldyrev} \&
  {Horaites}}{2019}]{boldyrev19}
{Boldyrev} S.,  {Horaites} K.,  2019, \mn@doi [\mnras] {10.1093/mnras/stz2378},
  \href {https://ui.adsabs.harvard.edu/abs/2019MNRAS.489.3412B} {489, 3412}

\bibitem[\protect\citeauthoryear{{Chandran}}{{Chandran}}{2010}]{chandran10}
{Chandran} B. D.~G.,  2010, \mn@doi [\apj] {10.1088/0004-637X/720/1/548}, \href
  {https://ui.adsabs.harvard.edu/abs/2010ApJ...720..548C} {720, 548}

\bibitem[\protect\citeauthoryear{{Chen}}{{Chen}}{2016}]{chen2016}
{Chen} C.~H.~K.,  2016, \mn@doi [Journal of Plasma Physics]
  {10.1017/S0022377816001124}, \href
  {http://adsabs.harvard.edu/abs/2016JPlPh..82f5302C} {82, 535820602}

\bibitem[\protect\citeauthoryear{{Cranmer}}{{Cranmer}}{2014}]{cranmer2014}
{Cranmer} S.~R.,  2014, \mn@doi [\apjs] {10.1088/0067-0049/213/1/16}, \href
  {https://ui.adsabs.harvard.edu/abs/2014ApJS..213...16C} {213, 16}

\bibitem[\protect\citeauthoryear{{Cranmer}, {van Ballegooijen}  \&
  {Edgar}}{{Cranmer} et~al.}{2007}]{cranmer2007}
{Cranmer} S.~R.,  {van Ballegooijen} A.~A.,   {Edgar} R.~J.,  2007, \mn@doi
  [\apjs] {10.1086/518001}, \href
  {https://ui.adsabs.harvard.edu/abs/2007ApJS..171..520C} {171, 520}

\bibitem[\protect\citeauthoryear{{Cranmer}, {Matthaeus}, {Breech}  \&
  {Kasper}}{{Cranmer} et~al.}{2009}]{cranmer2009}
{Cranmer} S.~R.,  {Matthaeus} W.~H.,  {Breech} B.~A.,   {Kasper} J.~C.,  2009,
  \mn@doi [\apj] {10.1088/0004-637X/702/2/1604}, \href
  {http://adsabs.harvard.edu/abs/2009ApJ...702.1604C} {702, 1604}

\bibitem[\protect\citeauthoryear{{Feldman}, {Asbridge}, {Bame}, {Montgomery}
  \& {Gary}}{{Feldman} et~al.}{1975}]{feldman75}
{Feldman} W.~C.,  {Asbridge} J.~R.,  {Bame} S.~J.,  {Montgomery} M.~D.,
  {Gary} S.~P.,  1975, \mn@doi [\jgr] {10.1029/JA080i031p04181}, \href
  {http://adsabs.harvard.edu/abs/1975JGR....80.4181F} {80, 4181}

\bibitem[\protect\citeauthoryear{{Gurevich} \& {Istomin}}{{Gurevich} \&
  {Istomin}}{1979}]{gurevichistomin79}
{Gurevich} A.~V.,  {Istomin} Y.~N.,  1979, Soviet Journal of Experimental and
  Theoretical Physics, \href
  {http://adsabs.harvard.edu/abs/1979JETP...50..470G} {50, 470}

\bibitem[\protect\citeauthoryear{{Helander} \& {Sigmar}}{{Helander} \&
  {Sigmar}}{2002}]{helandersigmar02}
{Helander} P.,  {Sigmar} D.~J.,  2002, {Collisional transport in magnetized
  plasmas. Cambridge University Press (Cambridge monographs on plasma
  physics;~4)}

\bibitem[\protect\citeauthoryear{{Horaites}, {Boldyrev}, {Krasheninnikov},
  {Salem}, {Bale}  \& {Pulupa}}{{Horaites} et~al.}{2015}]{horaites15}
{Horaites} K.,  {Boldyrev} S.,  {Krasheninnikov} S.~I.,  {Salem} C.,  {Bale}
  S.~D.,   {Pulupa} M.,  2015, \mn@doi [Physical Review Letters]
  {10.1103/PhysRevLett.114.245003}, \href
  {http://adsabs.harvard.edu/abs/2015PhRvL.114x5003H} {114, 245003}

\bibitem[\protect\citeauthoryear{{Horaites}, {Boldyrev}, {Wilson}, {Vi{\~n}as}
  \& {Merka}}{{Horaites} et~al.}{2018a}]{horaites18a}
{Horaites} K.,  {Boldyrev} S.,  {Wilson} III L.~B.,  {Vi{\~n}as} A.~F.,
  {Merka} J.,  2018a, \mn@doi [\mnras] {10.1093/mnras/stx2555}, \href
  {http://adsabs.harvard.edu/abs/2018MNRAS.474..115H} {474, 115}

\bibitem[\protect\citeauthoryear{{Horaites}, {Astfalk}, {Boldyrev}  \&
  {Jenko}}{{Horaites} et~al.}{2018b}]{horaites18b}
{Horaites} K.,  {Astfalk} P.,  {Boldyrev} S.,   {Jenko} F.,  2018b, \mn@doi
  [\mnras] {10.1093/mnras/sty1808}, \href
  {http://adsabs.harvard.edu/abs/2018MNRAS.480.1499H} {480, 1499}

\bibitem[\protect\citeauthoryear{{Horaites}, {Boldyrev}  \&
  {Medvedev}}{{Horaites} et~al.}{2019}]{horaites2019}
{Horaites} K.,  {Boldyrev} S.,   {Medvedev} M.~V.,  2019, \mn@doi [\mnras]
  {10.1093/mnras/sty3504}, \href
  {https://ui.adsabs.harvard.edu/abs/2019MNRAS.484.2474H} {484, 2474}

\bibitem[\protect\citeauthoryear{Huba, of Naval~Research  \& (U.S.)}{Huba
  et~al.}{1998}]{nrl}
Huba J.,  of Naval~Research U. S.~O.,   (U.S.) N. R.~L.,  1998, NRL Plasma
  Formulary.
NRL publication, Naval Research Laboratory, \url
  {http://books.google.com/books?id=h5CacQAACAAJ}

\bibitem[\protect\citeauthoryear{{Ivanov} \& {Prikhodko}}{{Ivanov} \&
  {Prikhodko}}{2013}]{ivanov2013}
{Ivanov} A.~A.,  {Prikhodko} V.~V.,  2013, \mn@doi [Plasma Physics and
  Controlled Fusion] {10.1088/0741-3335/55/6/063001}, \href
  {http://adsabs.harvard.edu/abs/2013PPCF...55f3001I} {55, 063001}

\bibitem[\protect\citeauthoryear{{Ivanov} \& {Prikhodko}}{{Ivanov} \&
  {Prikhodko}}{2017}]{ivanov2017}
{Ivanov} A.~A.,  {Prikhodko} V.~V.,  2017, \mn@doi [Physics Uspekhi]
  {10.3367/UFNe.2016.09.037967}, \href
  {http://adsabs.harvard.edu/abs/2017PhyU...60..509I} {60, 509}

\bibitem[\protect\citeauthoryear{{Klein} \& {Chandran}}{{Klein} \&
  {Chandran}}{2016}]{klein2016}
{Klein} K.~G.,  {Chandran} B. D.~G.,  2016, \mn@doi [\apj]
  {10.3847/0004-637X/820/1/47}, \href
  {https://ui.adsabs.harvard.edu/abs/2016ApJ...820...47K} {820, 47}

\bibitem[\protect\citeauthoryear{{K{\"o}hnlein}}{{K{\"o}hnlein}}{1996}]{koehnlein96}
{K{\"o}hnlein} W.,  1996, \mn@doi [\solphys] {10.1007/BF00153841}, \href
  {http://adsabs.harvard.edu/abs/1996SoPh..169..209K} {169, 209}

\bibitem[\protect\citeauthoryear{{Konkashbaev}, {Landman}  \&
  {Ulinich}}{{Konkashbaev} et~al.}{1978}]{konkashbaev1978}
{Konkashbaev} I.~K.,  {Landman} I.~S.,   {Ulinich} F.~R.,  1978, Soviet Journal
  of Experimental and Theoretical Physics, \href
  {http://adsabs.harvard.edu/abs/1978JETP...47..501K} {47, 501}

\bibitem[\protect\citeauthoryear{{Kulsrud}}{{Kulsrud}}{2005}]{kulsrud2005}
{Kulsrud} R.~M.,  2005, {Plasma physics for astrophysics}

\bibitem[\protect\citeauthoryear{{Landi}}{{Landi}}{2007}]{landi2007}
{Landi} E.,  2007, \mn@doi [\apj] {10.1086/517910}, \href
  {https://ui.adsabs.harvard.edu/abs/2007ApJ...663.1363L} {663, 1363}

\bibitem[\protect\citeauthoryear{{Landi} \& {Cranmer}}{{Landi} \&
  {Cranmer}}{2009}]{landi2009}
{Landi} E.,  {Cranmer} S.~R.,  2009, \mn@doi [\apj]
  {10.1088/0004-637X/691/1/794}, \href
  {https://ui.adsabs.harvard.edu/abs/2009ApJ...691..794L} {691, 794}

\bibitem[\protect\citeauthoryear{{L{\'o}pez}, {Shaaban}, {Lazar}, {Poedts},
  {Yoon}, {Micera}  \& {Lapenta}}{{L{\'o}pez} et~al.}{2019}]{lopez2019}
{L{\'o}pez} R.~A.,  {Shaaban} S.~M.,  {Lazar} M.,  {Poedts} S.,  {Yoon} P.~H.,
  {Micera} A.,   {Lapenta} G.,  2019, \mn@doi [\apjl]
  {10.3847/2041-8213/ab398b}, \href
  {https://ui.adsabs.harvard.edu/abs/2019ApJ...882L...8L} {882, L8}

\bibitem[\protect\citeauthoryear{{Mercier} \& {Chambe}}{{Mercier} \&
  {Chambe}}{2015}]{mercier2015}
{Mercier} C.,  {Chambe} G.,  2015, \mn@doi [\aap]
  {10.1051/0004-6361/201425540}, \href
  {https://ui.adsabs.harvard.edu/abs/2015A%26A...583A.101M} {583, A101}

\bibitem[\protect\citeauthoryear{{Mirnov} \& {Ryutov}}{{Mirnov} \&
  {Ryutov}}{1979}]{mirnov1979}
{Mirnov} V.~V.,  {Ryutov} D.~D.,  1979, Technical Physics Letters, \href
  {http://adsabs.harvard.edu/abs/1979TePhL...5..279M} {5, 678}

\bibitem[\protect\citeauthoryear{{Noci}}{{Noci}}{2003}]{noci2003}
{Noci} G.,  2003, \memsai, \href
  {https://ui.adsabs.harvard.edu/abs/2003MmSAI..74..704N} {74, 704}

\bibitem[\protect\citeauthoryear{{Perkins}}{{Perkins}}{1973}]{perkins1973}
{Perkins} F.,  1973, \mn@doi [\apj] {10.1086/151902}, \href
  {https://ui.adsabs.harvard.edu/abs/1973ApJ...179..637P} {179, 637}

\bibitem[\protect\citeauthoryear{{Pierrard}, {Lazar}, {Poedts}, {{\v
  S}tver{\'a}k}, {Maksimovic}  \& {Tr{\'a}vn{\'{\i}}{\v c}ek}}{{Pierrard}
  et~al.}{2016}]{pierrard2016}
{Pierrard} V.,  {Lazar} M.,  {Poedts} S.,  {{\v S}tver{\'a}k} {\v S}.,
  {Maksimovic} M.,   {Tr{\'a}vn{\'{\i}}{\v c}ek} P.~M.,  2016, \mn@doi
  [\solphys] {10.1007/s11207-016-0961-7}, \href
  {http://adsabs.harvard.edu/abs/2016SoPh..291.2165P} {291, 2165}

\bibitem[\protect\citeauthoryear{{Pilipp}, {Muehlhaeuser}, {Miggenrieder},
  {Montgomery}  \& {Rosenbauer}}{{Pilipp} et~al.}{1987}]{pilipp87}
{Pilipp} W.~G.,  {Muehlhaeuser} K.-H.,  {Miggenrieder} H.,  {Montgomery} M.~D.,
    {Rosenbauer} H.,  1987, \mn@doi [\jgr] {10.1029/JA092iA02p01075}, \href
  {http://adsabs.harvard.edu/abs/1987JGR....92.1075P} {92, 1075}

\bibitem[\protect\citeauthoryear{{Post}}{{Post}}{1987}]{post1987}
{Post} R.~F.,  1987, Nucl. Fusion, 27, 1579

\bibitem[\protect\citeauthoryear{{Richardson} \& {Smith}}{{Richardson} \&
  {Smith}}{2003}]{richardson2003}
{Richardson} J.~D.,  {Smith} C.~W.,  2003, \mn@doi [\grl]
  {10.1029/2002GL016551}, \href
  {https://ui.adsabs.harvard.edu/abs/2003GeoRL..30.1206R} {30, 1206}

\bibitem[\protect\citeauthoryear{{Roberg-Clark}, {Agapitov}, {Drake}  \&
  {Swisdak}}{{Roberg-Clark} et~al.}{2019}]{RC2019}
{Roberg-Clark} G.~T.,  {Agapitov} O.,  {Drake} J.~F.,   {Swisdak} M.,  2019,
  \mn@doi [\apj] {10.3847/1538-4357/ab5114}, \href
  {https://ui.adsabs.harvard.edu/abs/2019ApJ...887..190R} {887, 190}

\bibitem[\protect\citeauthoryear{{Ryutov}}{{Ryutov}}{2005}]{ryutov2005}
{Ryutov} D.~D.,  2005, Fusion Sci. Technol., 47, 148

\bibitem[\protect\citeauthoryear{{Scudder} \& {Olbert}}{{Scudder} \&
  {Olbert}}{1979}]{scudderolbert79}
{Scudder} J.~D.,  {Olbert} S.,  1979, \mn@doi [\jgr] {10.1029/JA084iA11p06603},
  \href {http://adsabs.harvard.edu/abs/1979JGR....84.6603S} {84, 6603}

\bibitem[\protect\citeauthoryear{{Shaaban}, {Lazar}, {Yoon}, {Poedts}  \&
  {L{\'o}pez}}{{Shaaban} et~al.}{2019}]{shaaban2019}
{Shaaban} S.~M.,  {Lazar} M.,  {Yoon} P.~H.,  {Poedts} S.,   {L{\'o}pez} R.~A.,
   2019, \mn@doi [\mnras] {10.1093/mnras/stz830}, \href
  {https://ui.adsabs.harvard.edu/abs/2019MNRAS.486.4498S} {486, 4498}

\bibitem[\protect\citeauthoryear{{Spitzer} \& {H{\"a}rm}}{{Spitzer} \&
  {H{\"a}rm}}{1953}]{spitzerharm53}
{Spitzer} L.,  {H{\"a}rm} R.,  1953, \mn@doi [Physical Review]
  {10.1103/PhysRev.89.977}, \href
  {http://adsabs.harvard.edu/abs/1953PhRv...89..977S} {89, 977}

\bibitem[\protect\citeauthoryear{{Stawarz}, {Smith}, {Vasquez}, {Forman}  \&
  {MacBride}}{{Stawarz} et~al.}{2009}]{stawarz2009}
{Stawarz} J.~E.,  {Smith} C.~W.,  {Vasquez} B.~J.,  {Forman} M.~A.,
  {MacBride} B.~T.,  2009, \mn@doi [\apj] {10.1088/0004-637X/697/2/1119}, \href
  {https://ui.adsabs.harvard.edu/abs/2009ApJ...697.1119S} {697, 1119}

\bibitem[\protect\citeauthoryear{{Tang}, {Zank}  \& {Kolobov}}{{Tang}
  et~al.}{2018}]{tang2018}
{Tang} B.,  {Zank} G.~P.,   {Kolobov} V.,  2018, in Journal of Physics
  Conference Series. p. 012025, \mn@doi{10.1088/1742-6596/1100/1/012025}

\bibitem[\protect\citeauthoryear{{Vasko}, {Krasnoselskikh}, {Tong}, {Bale},
  {Bonnell}  \& {Mozer}}{{Vasko} et~al.}{2019}]{vasko2019}
{Vasko} I.~Y.,  {Krasnoselskikh} V.,  {Tong} Y.,  {Bale} S.~D.,  {Bonnell}
  J.~W.,   {Mozer} F.~S.,  2019, \mn@doi [\apjl] {10.3847/2041-8213/ab01bd},
  \href {https://ui.adsabs.harvard.edu/abs/2019ApJ...871L..29V} {871, L29}

\bibitem[\protect\citeauthoryear{{Vech}, {Klein}  \& {Kasper}}{{Vech}
  et~al.}{2017}]{vech2017}
{Vech} D.,  {Klein} K.~G.,   {Kasper} J.~C.,  2017, \mn@doi [\apjl]
  {10.3847/2041-8213/aa9887}, \href
  {https://ui.adsabs.harvard.edu/abs/2017ApJ...850L..11V} {850, L11}

\bibitem[\protect\citeauthoryear{{Verscharen}, {Klein}  \&
  {Maruca}}{{Verscharen} et~al.}{2019a}]{verscharen2019}
{Verscharen} D.,  {Klein} K.~G.,   {Maruca} B.~A.,  2019a, arXiv e-prints:
  1902.03448, \href {http://adsabs.harvard.edu/abs/2019arXiv190203448V} {}

\bibitem[\protect\citeauthoryear{{Verscharen}, {Chandran}, {Jeong}, {Salem},
  {Pulupa}  \& {Bale}}{{Verscharen} et~al.}{2019b}]{verscharen2019a}
{Verscharen} D.,  {Chandran} B. D.~G.,  {Jeong} S.-Y.,  {Salem} C.~S.,
  {Pulupa} M.~P.,   {Bale} S.~D.,  2019b, \mn@doi [\apj]
  {10.3847/1538-4357/ab4c30}, \href
  {https://ui.adsabs.harvard.edu/abs/2019ApJ...886..136V} {886, 136}

\bibitem[\protect\citeauthoryear{{{\v S}tver{\'a}k}, {Tr{\'a}vn{\'{\i}}{\v
  c}ek}  \& {Hellinger}}{{{\v S}tver{\'a}k} et~al.}{2015}]{stverak15}
{{\v S}tver{\'a}k} {\v S}.,  {Tr{\'a}vn{\'{\i}}{\v c}ek} P.~M.,   {Hellinger}
  P.,  2015, \mn@doi [Journal of Geophysical Research (Space Physics)]
  {10.1002/2015JA021368}, \href
  {http://adsabs.harvard.edu/abs/2015JGRA..120.8177A} {120, 8177}

\makeatother
\end{thebibliography}










\end{document}